\documentclass[runningheads]{llncs}
\usepackage{graphicx}
\usepackage{mathtools}
\usepackage{amsmath}
\usepackage{amsfonts}
\usepackage{booktabs}
\usepackage{dirtytalk}
\begin{document}
%
\title{Unsupervised Anomaly Localization with Structural Feature-Autoencoders}
\titlerunning{Structural Feature-Autoencoders}
%
\author{Felix Meissen\inst{1,2} \and
Johannes Paetzold\inst{1,2} \and
Georgios Kaissis\inst{1,2,3} \and
Daniel Rueckert\inst{1,2,3}}
\authorrunning{F. Meissen et al.}
%
\institute{Technical University of Munich (TUM), Munich, Germany \and
Klinikum Rechts der Isar, Munich, Germany \and
Imperial College London, UK
\email{\{felix.meissen,g.kaissis,daniel.rueckert\}@tum.de}}
%
\maketitle              
\begin{abstract}
Unsupervised Anomaly Detection has become a popular method to detect pathologies in medical images as it does not require supervision or labels for training.
Most commonly, the anomaly detection model generates a \say{normal} version of an input image, and the pixel-wise $l^p$-difference of the two is used to localize anomalies.
However, large residuals often occur due to imperfect reconstruction of the complex anatomical structures present in most medical images.
This method also fails to detect anomalies that are not characterized by large intensity differences to the surrounding tissue.
We propose to tackle this problem using a feature-mapping function that transforms the input intensity images into a space with multiple channels where anomalies can be detected along different discriminative feature maps extracted from the original image.
We then train an Autoencoder model in this space using structural similarity loss that does not only consider differences in intensity but also in contrast and structure.
Our method significantly increases performance on two medical data sets for brain MRI.
Code and experiments are available at \url{https://github.com/FeliMe/feature-autoencoder}

\keywords{Semi-Supervised Learning \and Anomaly Localization \and Anomaly Detection}
\end{abstract}
\section{Introduction}

\begin{figure}[htpb]
    \centering
    \includegraphics[width=1.0\textwidth]{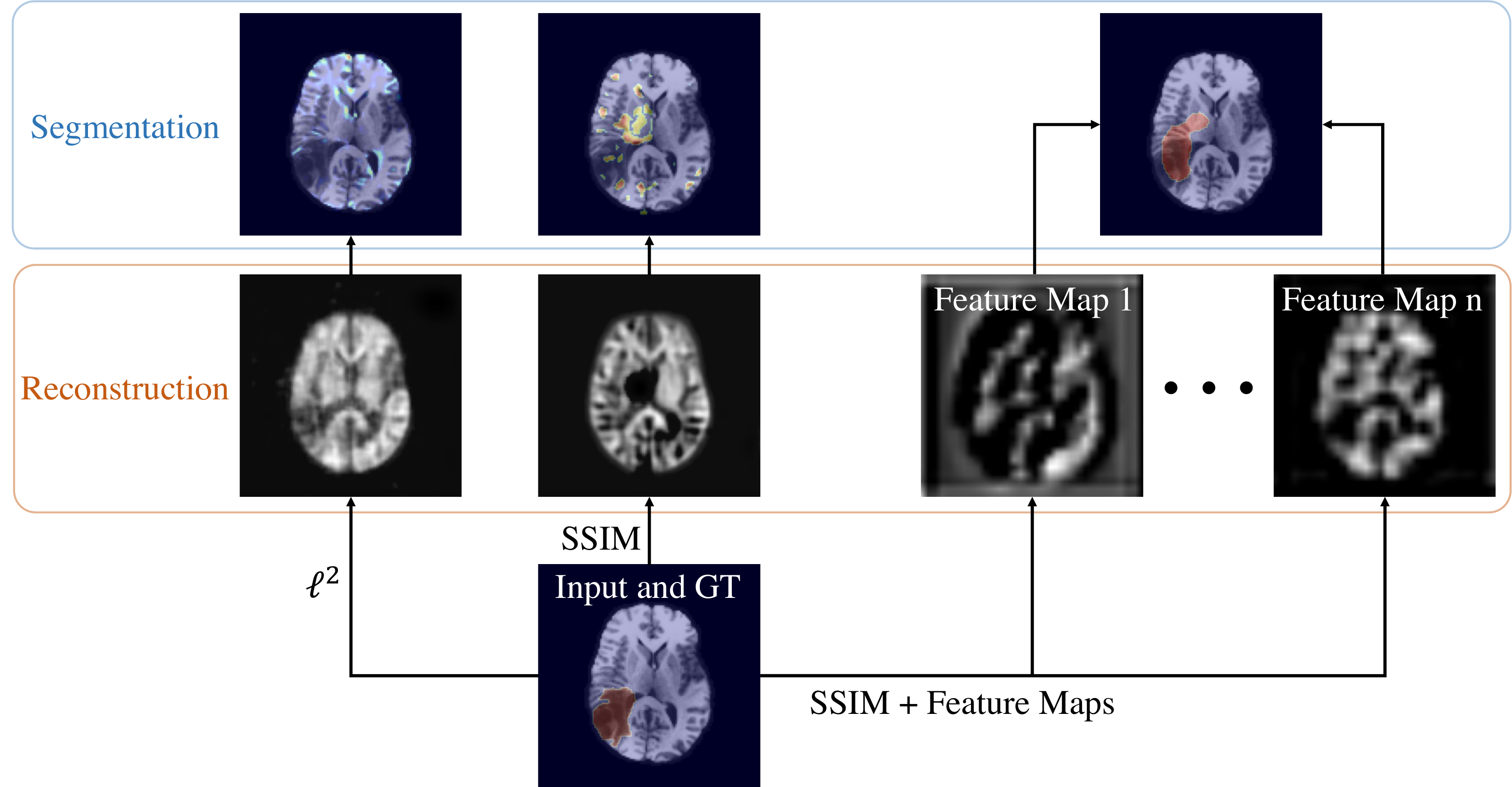}
    \caption{A sample of a brain MR image showing a tumor (bottom) is reconstructed by a model trained to minimize the $L_2$-loss (left) SSIM-loss (middle) and to reconstruct feature maps in a multi-channel feature space with SSIM-loss (right). The combination of $n$ anomaly maps gives the most accurate and focused localization.}
    \label{fig:demo}
\end{figure}

For computed aided decision support systems, accurate detection and localization of pathologies is crucial to assist the radiologist in their work and to build trust in the underlying machine learning algorithms. Supervised methods require large amounts of labeled data. However, manual labeling of medical images -- especially pixel- or voxel-wise segmentation -- is expensive, time-consuming, and can be ambiguous if two raters don't agree on the same contour of a certain pathology.
Unsupervised anomaly localization algorithms can detect regions in images that deviate from the normal appearance without ever seeing anomalous samples of any class during training.
These algorithms use machine learning models to learn the normative distribution of the data from normal samples only.
Generative machine learning models, such as Autoencoders (AE) or Generative Adversarial Networks (GAN), have been proven successful for this task.
For a new input image, these models usually generate a reconstruction that lies within the distribution of the \say{normal} data and detect anomalies from pixel-wise residual maps between the input- and the reconstructed image.
In \cite{anogan} and \cite{fanogan}, Schlegl \textit{et al.} apply this principle by training a GAN on images of retinal OCT scans. They generate the reconstruction during inference by mapping the new sample to the latent space of the GAN.
In \cite{CTAE}, Pawlowski \textit{et al.} use an Autoencoder to detect anomalies in CT images of the brain.
Zimmerer \textit{et al.} \cite{BrainVAE} train a Variational Autoencoder (VAE) to maximize the likelihood of the normal training data.
They experiment with different anomaly scoring functions and found the gradient of the evidence lower bound (ELBO) with respect to the pixels of the input image to improve detection performance.
In \cite{ComparativeStudy}, Baur \textit{et al.} present a thorough comparison of different anomaly localization approaches.
However, for detecting tumors and lesions in brain MRI, Saase \textit{et al.} and Meissen \textit{et al.} have independently shown that these methods can be outperformed by simple statistical methods \cite{saase,meissen1}.
In later work, Meissen \textit{et al.} have identified that small reconstruction inaccuracies at edges and complex anatomical structures yield large residuals that hinder the detection of anomalies that are not characterized by high-contrast intensity regions \cite{meissen2}. They also hypothesize that this problem is more pronounced for single-channel images than for multi-channel images because, in the latter one, the anomaly-contributions of different channels to each pixel add up.
To alleviate these problems, we apply two changes to the standard Autoencoder framework that were successful in industrial defect detection.
Similar to DFR \cite{dfr}, we employ a feature-mapping function to transform the data into a multi-channel space, so that deviations from the distribution of healthy data can be detected along multiple feature maps.
We use structural similarity (SSIM) \cite{SSIMAE,ssim} to train our network and localize anomalies. SSIM captures dissimilarities in multiple ways, including structural and contrast differences.
Our contributions are the following:

\begin{itemize}
    \item We combine working in a multi-channel feature space with SSIM for anomaly localization.
    \item We propose a specialized architecture design which is better suited for training with SSIM in feature spaces.
    \item We show that combining these improvements leads to significant performance gains, by evaluating our method on two data sets of medical images where anomalies do not appear as hyperintense regions.
\end{itemize}

Our novel approach outperforms all comparing methods in our experiments.
Figure \ref{fig:demo} shows the effectiveness of our proposed Structural Feature-Autoencoder on a sample from the Multimodal Brain Tumor Image Segmentation Benchmark (BraTS) data set \cite{brats1,brats2,brats3}.

\section{Methodology} \label{sec:method}

\begin{figure}[htpb]
    \centering
    \includegraphics[width=1.0\textwidth]{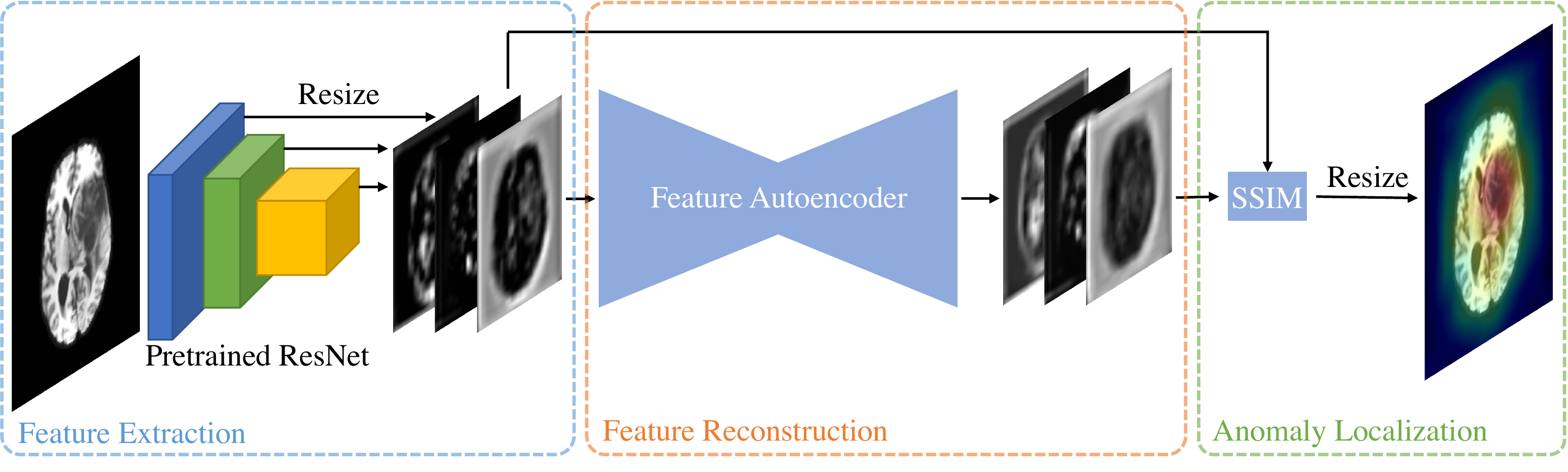}
    \caption{Pipeline of our proposed method: First, we extract image feature maps with a pre-trained CNN, then we reconstruct these features using an Autoencoder. Finally, we localize anomalies via SSIM between the extracted and the reconstructed feature maps.}
    \label{fig:pipeline}
\end{figure}

We train a vanilla convolutional Autoencoder to reconstruct feature maps extracted from the input images as shown in Figure \ref{fig:pipeline}.

\subsection{Feature Extraction} \label{sec:feature_extraction}

We use a pre-trained CNN to transform our initial input image $x \in \mathbb{R}^{c \times h \times w}$ into a multi-channel space with discriminative hierarchical features.
The CNN can be seen as a combination of $L$ sequential feature mapping functions, each of them outputting a feature map $\phi_l(\mathbf{x})$ with size $c_l \times h_l \times w_l$.
While early layers with small receptive fields capture local facets of the input image, later layers encode more global information.
Our extractor combines feature maps from different scales by resizing all maps $\phi_l(\mathbf{x})$ to the spatial size of the largest feature map $\phi_1(\mathbf{x})$ using bilinear interpolation.

\subsection{SSIM Loss}

The Structural Similarity Index (SSIM) \cite{ssim} is a metric to measure the perceptual similarity of two images by comparing them on a patch level.
For every patch $\mathbf{p}$ in the first image and their corresponding patch $\mathbf{q}$ in the second image, SSIM computes their similarity in terms of luminance $l(\mathbf{p}, \mathbf{q})$, contrast $c(\mathbf{p}, \mathbf{q})$, and structure $s(\mathbf{p}, \mathbf{q})$, and aggregates them into a single score.
Similarity in luminance is calculated by comparing the means of the two patches $\mu_\mathbf{p}$ and $\mu_\mathbf{q}$, contrast by comparing the variances $\sigma_\mathbf{p}$ and $\sigma_\mathbf{q}$, and for structure, the covariance $\sigma_{\mathbf{p}\mathbf{q}}$ is used.
Together with two constants $C_1$ and $C_2$ for numerical stability, SSIM equates to:

\begin{equation*}
    \text{SSIM}(\mathbf{p}, \mathbf{q}) = l(\mathbf{p}, \mathbf{q}) c(\mathbf{p}, \mathbf{q}) s(\mathbf{p}, \mathbf{q}) = \frac{\left( 2 \mu_\mathbf{p} \mu_\mathbf{q} + C_1 \right) \left( 2 \sigma_{\mathbf{p}\mathbf{q}} + C_2 \right)}{\left( \mu_\mathbf{p}^2 + \mu_\mathbf{q}^2 + C_1 \right) \left( \sigma_\mathbf{p}^2 + \sigma_\mathbf{q}^2 + C_2 \right)}
\end{equation*}

By computing a score for every pixel-location in the input- and reconstructed image $\mathbf{x}$ and $\mathbf{\hat{x}} \in \mathbb{R}^{c \times h \times w}$, the algorithm outputs an anomaly map $\mathbf{a} \in \mathbb{R}^{h \times w}$, or a scalar loss term when computing the mean SSIM (MSSIM) that evaluates the overall similarity and is differentiable.

\begin{equation*}
    \text{MSSIM}(\mathbf{x}, \mathbf{\hat{x}}) = \frac{1}{M} \sum^M_{j=1} \text{SSIM}(\mathbf{x}_j, \mathbf{\hat{x}}_j)
\end{equation*}

$M = h \times w$.
We use MSSIM as a loss function to train our model and the anomaly maps $\mathbf{a}$ to localize anomalies in the image.

\section{Experiments} \label{sec:experiments}

\subsection{Datasets}

We evaluate our method on two publicly available data sets of brain MRI.
First, we use the data of the MOOD Analysis Challenge 2020 \cite{mood}.
It consists of 800 T2-weighted scans of healthy young adults from which we use 640 for training and split the remaining into $10\%$ validation and $90\%$ test.
As part of the pre-processing, we perform histogram equalization on every scan and use a subset of 80 slices around the center of the brain because the majority of lesions are usually accumulated in this region.
We then resize all slices to $128 \times 128$ pixels and add an artificial sink deformation anomaly from FPI \cite{fpi} to half of the resulting images in the validation and test sets.
Note that the sink deformation anomaly has similar pixel intensities as the surrounding tissue and can therefore not easily be detected via thresholding \cite{meissen2}.
Apart from that, the MOOD evaluation data does not contain any anomalies.

The second evaluation is performed on a real world data set.
Here, we use the Cambridge Centre for Ageing and Neuroscience (Cam-CAN) dataset \cite{camcan} for training.
It contains scans from 653 healthy women and men between the age of 18 and 87.
Both reported genders and all age groups are approximately uniformly distributed.
All scans were acquired with a 3T Siemens Magnetom TrioTim syngo MR machine at the Medical Research Council Cognition and Brain Sciences Unit in Cambridge, UK.
For evaluation, we use the training set of the 2020 version of the Multimodal Brain Tumor Image Segmentation Benchmark (BraTS) \cite{brats2,brats3,brats1}.
Here, scans of 371 patients were acquired with different clinical protocols and various scanners from 19 institutions.
The BraTS dataset has manual segmentations from up to four raters.
36 of the BraTS-scans are used for validation, the remainder is used for testing.
We only use T1-weighted scans for training and evaluation of the BraTS dataset to assess the detection performance of our method on anomalies that are not necessarily hyperintense.
All scans in Cam-CAN and BraTS are registered to the SRI atlas \cite{sri} and the scans in Cam-CAN are additionally skull-stripped using ROBEX \cite{robex} beforehand.
We perform the same pre-processing as for the MOOD data.
It is important to note that the distributions of the Cam-CAN training and BraTS evaluation data are likely to be different because of varying scanner types and protocols in the respective acquisition sites.
This poses difficulties for Unsupervised Anomaly Detection models since they are designed to detect out-of-distribution samples and even slight distributions shifts might be picked up by the them.

\subsection{Implementation}

As feature-mapping function, we choose a ResNet18 \cite{resnet} pre-trained on ImageNet \cite{imagenet}.
If not specified otherwise, we use the feature maps of the first three layers (layer0, layer1, and layer2) with a spatial resolution of $32 \times 32$, $32 \times 32$, and $16 \times 16$ respectively.
The Feature-Autoencoder is a fully convolutional Autoencoder with four layers in the encoder and the decoder.
All encoder layers consist of a $5 \times 5$ convolution with stride $2$, same padding, and without bias, followed by a batch normalization layer, leaky ReLU activations with a negative slope of $0.01$, and dropout with a probability of $0.1$.
This contrasty DFR which only uses $1 \times 1$ convolutions and no spatial down- and upsampling.
The encoder is wrapped up by a convolution with a $5 \times 5$ kernel and stride 1.
The numbers of channels for the encoder layers are $100$, $150$, $200$, and $300$.
The decoder mirrors the encoder in the number of channels and has four layers with strided transpose convolutions for upsampling, followed by batch normalization, leaky ReLU activations, and dropout as in the encoder.
The decoder additionally has a final $1$ by $1$ convolution which outputs the reconstructed feature maps.
We implement our model in PyTorch \cite{pytorch} and train it using the Adam optimizer \cite{adam} with a learning rate of $0.0002$, and a batch size of 64 for $10.000$ steps.

\subsection{Baselines} \label{sec:baselines}

We compare our method against several baselines for semi- and self-supervised anomaly localization:
We choose the VAE proposed by Zimmerer \textit{et al.} \cite{BrainVAE} as a method that uses likelihood-measures to detect anomalies. We use the best-performing methods from their paper (\say{combi} for localization and the KL-term for detection).
For image reconstruction-based variants, we compare against a Vanilla AE with $L_2$-loss (AE MSE) and one with SSIM-loss (AE SSIM) \cite{SSIMAE}.
We select the same architectures as for the VAE but exchange the variational bottleneck.
Results are also compared against those of f-AnoGAN \cite{fanogan}.
We further compare against DFR \cite{dfr} and DFR trained with SSIM-loss to quantify the influence of our architectural contribution.
Next, we use the recently proposed self-supervised methods Foreign Patch Interpolation (FPI) \cite{fpi} and Poisson Image Interpolation (PII) \cite{pii} that train a segmentation model on synthetic anomalies.
Lastly, we evaluate the statistical baseline method (BM) from Saase \textit{et al.} \cite{saase} that previously outperformed several reconstruction-based methods.
For all baselines that require training, we choose Adam with a learning rate of $0.0002$ as an optimizer, set the batch size to $64$, and train for $10.000$ steps.
Only DFR uses the original batch size of $4$ from the paper, because of its large memory footprint.

\subsection{Evaluation metrics}

For evaluating anomaly localization performance of the models, we resort to standard metrics used in the literature.
First, we use the pixel-wise average precision (Pixel-AP) which is equivalent to the area under the precision-recall curve and is threshold-independent.
We also report the Sørensen–Dice coefficient at a false positive rate of $5\%$ (Dice @ 5\% FPR).
Sample-wise performance is measured via the area under the receiver operating characteristics curve (Image-AUROC).
We considered every slice that contains at least one anomalous pixel as abnormal.
All metrics are computed over the whole test data set.

\section{Results} \label{sec:results}

\begin{table}[htpb]
\caption{Performance of all compared models and a random classifier, including standard deviations over $N=5$ runs with different random seeds, on the MOOD data set with sink deformation anomalies.}
\label{tab:results_mood}
\centering
\bgroup
\begin{tabular}{llll}
\toprule
                        & Pixel-AP          & Dice @ $5\%$ FPR  & Image-AUROC   \\
\cmidrule(r{4pt}){1-1} \cmidrule(){2-4}
Random                  & $0.019 \pm 0.000$ & $0.028 \pm 0.000$ & $0.500 \pm 0.000$ \\
VAE \cite{BrainVAE}     & $0.051 \pm 0.000$ & $0.092 \pm 0.001$ & $0.482 \pm 0.003$ \\
AE MSE                  & $0.048 \pm 0.001$ & $0.088 \pm 0.001$ & $0.572 \pm 0.006$ \\
AE SSIM \cite{SSIMAE}   & $0.090 \pm 0.010$ & $0.155 \pm 0.022$ & $0.624 \pm 0.004$ \\
f-AnoGAN \cite{fanogan} & $0.052 \pm 0.001$ & $0.071 \pm 0.001$ & $0.580 \pm 0.007$ \\
FPI \cite{fpi}          & $\mathbf{0.474 \pm 0.048}$ & $0.320 \pm 0.017$ & $\mathbf{0.879 \pm 0.046}$ \\
PII \cite{pii}          & $0.338 \pm 0.146$ & $0.268 \pm 0.064$ & $0.808 \pm 0.078$ \\
DFR \cite{dfr}          & $0.080 \pm 0.001$ & $0.138 \pm 0.003$ & $0.560 \pm 0.004$ \\
DFR SSIM                & $0.080 \pm 0.002$ & $0.140 \pm 0.003$ & $0.556 \pm 0.004$ \\
BM \cite{saase}         & $0.048 \pm 0.000$ & $0.090 \pm 0.000$ & $0.544 \pm 0.000$ \\
\cmidrule(r{4pt}){1-1} \cmidrule(){2-4}
\textbf{Ours}           & $0.431 \pm 0.007$ & $\mathbf{0.336 \pm 0.004}$ & $0.775 \pm 0.007$ \\
\bottomrule
\end{tabular}
\egroup
\end{table}

\begin{figure}[htpb]
    \centering
    \includegraphics[width=1.0\textwidth]{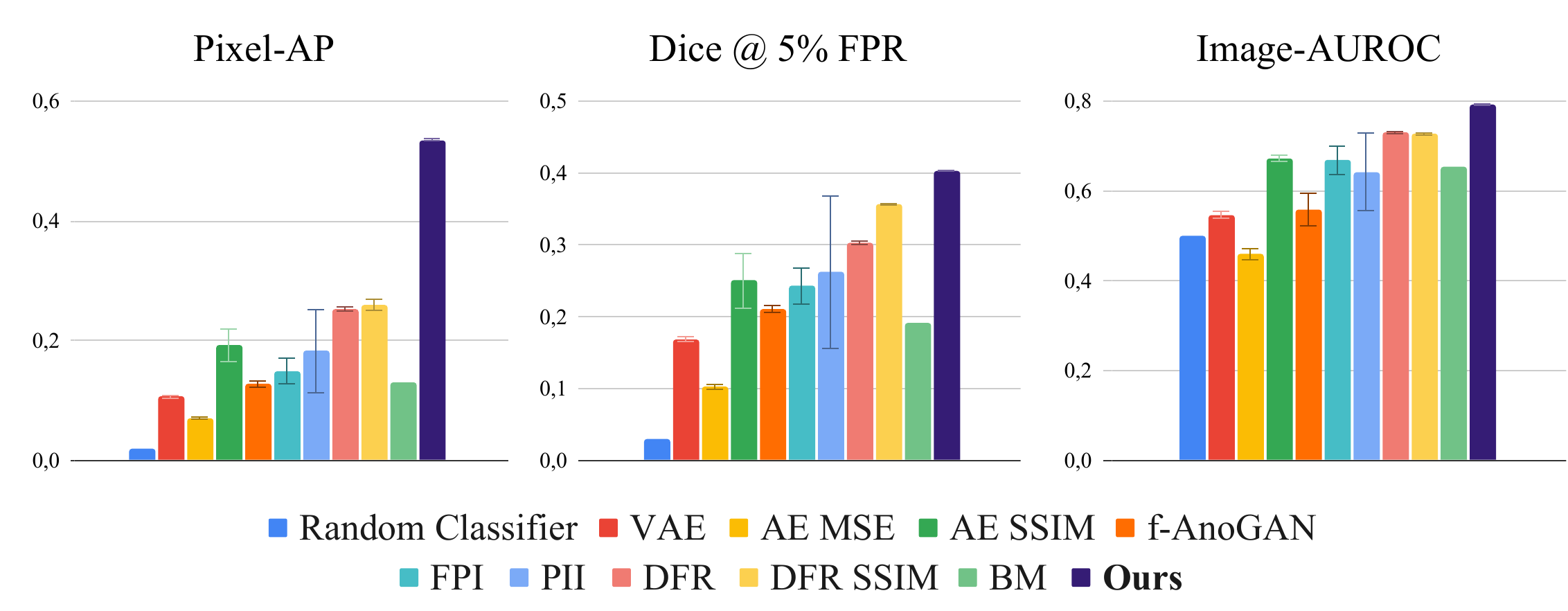}
    \caption{Performance of all compared models and a random classifier, including error bars that indicate one standard deviation over $N=5$ runs with different random seeds, on the BraTS data set. Our method performs significantly better than all compared methods (t-test; $p \leq 0.05$).}
    \label{fig:results_brats}
\end{figure}

We repeat the training- and evaluation procedure for each method with $N=5$ different random seeds. The results for the MOOD and BraTS data sets are shown in Table \ref{tab:results_mood} and Figure \ref{fig:results_brats} respectively.
On MOOD, only the self-supervised methods FPI outperform our proposed model.
However, it is crucial to note that at FPI was designed to achieve good performance on this anomaly type.
For the real-world data set, our model outperforms all competing methods from Section \ref{sec:baselines} both in the pixel- and in the sample-wise setting.
We perform a two-sided, heteroscedastic t-test with $p \leq 0.05$ to show that the results are statistically significant.
We also added the performance a random uniform classifier would get (Random Classifier) for comparison.

\subsection{Ablation - Feature Extractor Layers}

\begin{figure}[htpb]
    \centering
    \includegraphics[width=1.0\textwidth]{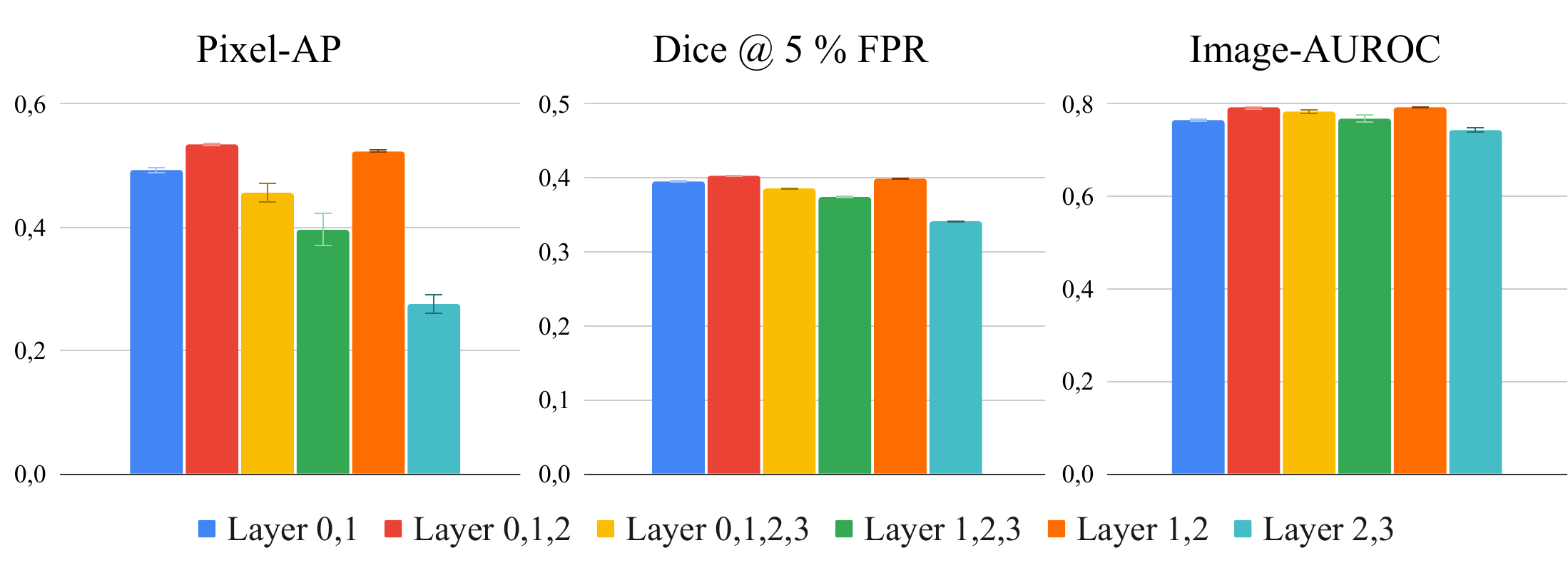}
    \caption{Ablation study using different layers of the feature-mapping function. The error bars indicate one standard deviation.}
    \label{fig:results_ablation}
\end{figure}

We perform an ablation study to evaluate the influence of the feature-mapping function on the performance of our method. We use feature maps from different layers of the pre-trained ResNet (Indicated by \say{layer x,y,z}). The results are shown in Figure \ref{fig:results_ablation}.

\section{Discussion} \label{sec:discussion}

The above experiments provide evidence that our contributions significantly improve anomaly detection and localization performance on anomalies that are mainly characterized by textural differences.
Although the anomalies are not hyperintense and despite the domain shift between training and evaluation data, our model achieves strong performance on both data sets.

\subsection{Discussion of the Main Experiments}

The results of our method and DFR have low variance compared to others, which demonstrates that anomaly localization in a multi-channel feature space is more robust than in image space.
The two self-supervised methods perform well in detecting the artificial sink anomalies.
This is not surprising as they have been designed and optimized for this exact type of artifical anomaly.
Importantly, however, they fail in detecting real-world anomalies, which indicates that the self-supervision task does not generalize beyond synthetic anomalies.
They also consistently exhibit the largest variance in the pixel- and the sample-wise task.
Their performance, therefore, is more random and less reliable than our method.
Noticeably, DFR with SSIM-loss performs worse than our method and only slightly better than vanilla DFR (Figure \ref{fig:results_brats}).
This is a result of the architecture choices of both methods:
DFR uses only 1 by 1 convolutions with no spatial interconnections.
However, a scoring function like SSIM -- that works on patches -- benefits from an architecture like ours with a large receptive field provided by spatial convolutions, and down- and upsampling.
SSIM captures inter-dependencies between neighboring pixels and combines differences along three dimensions (luminance, contrast, and structure), while the pixel-wise residual scores pixels isolated and only based on their intensity.
Combining this process on multiple channels multiplies the beneficial effect.

\subsection{Discussion of the Ablation}

The ablation experiment shows a consistent ranking for both pixel- and sample-wise performance of the feature-extractor layers.
Earlier layers of the ResNet seem to provide more useful features than later layers with layers 1 and 2 being the most important ones.
This makes sense as CNN-based backbones are known to extract a hierarchy of features with lower layers containing more textural and later ones more semantic information.
On the other, a larger depth of the feature maps is harder to fit as it requires more parameters and, thus, more training samples. That is why a drop in performance between \say{layer 0, 1, 2} and \say{layer 0, 1, 2, 3} can be observed.

\subsection{Limitations}

\begin{figure}[htpb]
    \centering
    \includegraphics[width=1.0\textwidth]{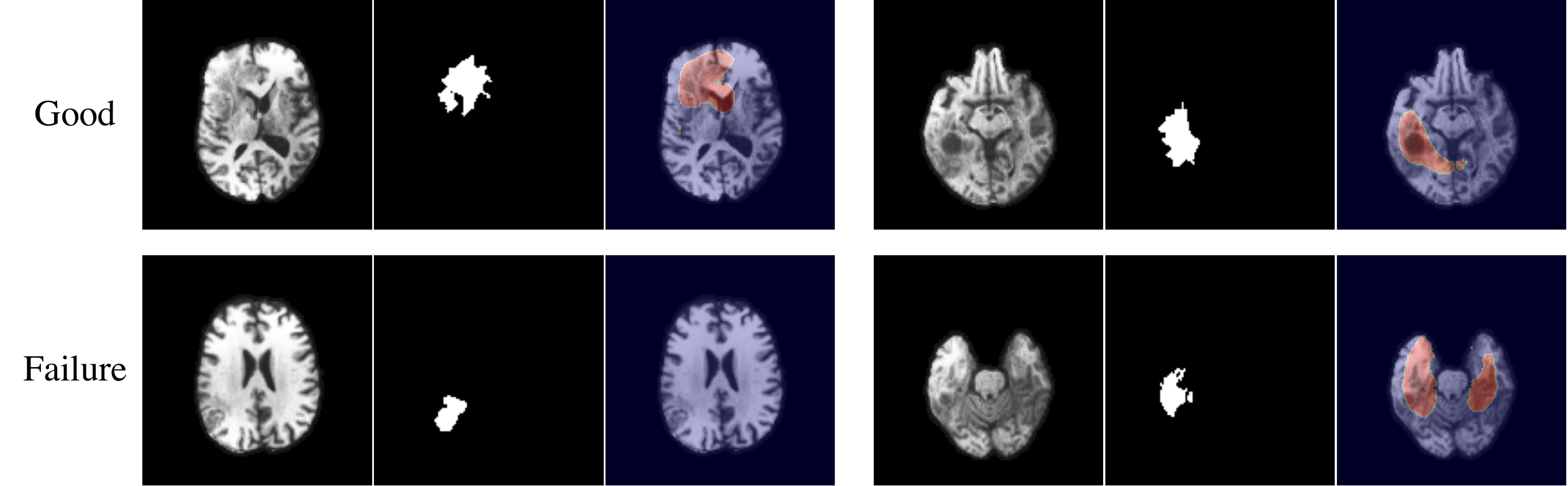}
    \caption{Two successful samples (first row) and two failure cases (second row). On the left, the model fails to detect a small anomaly. On the right, a slice far away from the center shows false positives. Each with input image, manual segmentation, and anomaly map thresholded at $t=0.75$ from left to right.}
    \label{fig:good_and_bad}
\end{figure}

While working in multi-channel feature spaces improves robustness and performance of anomaly detection methods for brain MRI, the loss of spatial resolution through the feature mapping function leads to coarse segmentations and can cause models to miss small anomalies as shown in Figure \ref{fig:good_and_bad}. Another failure case shown here happens at slices far away from the center of the brain. Since the structure there differs significantly from the rest of the brain, large regions get misclassified as anomalous.

\section{Conclusion}

This work has shown that using SSIM in a multi-channel discriminative feature space improves localization of non-hyperintense anomalies on brain MRI significantly and sets a new state-of-the-art in this domain.
While we are aware that for unsupervised detection of brain tumors other MR sequences, such as fluid-attenuated inversion recovery (FLAIR) images are available, we argue that our method can easily be transferred to other modalities and pathologies, such as CT, where this isn't the case.
We deem anomaly detection from feature spaces a promising research direction and will further explore methods to increase the spatial resolution of these approaches.
We will also apply our findings to different modalities and pathologies and study their behavior.

%
%
%
%
\bibliographystyle{splncs04}
\bibliography{references}

\begin{thebibliography}{10}
\providecommand{\url}[1]{\texttt{#1}}
\providecommand{\urlprefix}{URL }
\providecommand{\doi}[1]{https://doi.org/#1}

\bibitem{brats2}
Bakas, S., Akbari, H., Sotiras, A., Bilello, M., Rozycki, M., Kirby, J.S.,
  Freymann, J.B., Farahani, K., Davatzikos, C.: Advancing the cancer genome
  atlas glioma mri collections with expert segmentation labels and radiomic
  features. Scientific Data  \textbf{4},  2052--4463 (2017).
  \doi{10.1038/sdata.2017.117}

\bibitem{brats3}
Bakas, S., Reyes, M., Jakab, A., Bauer, S., Rempfler, M., Crimi, A., Shinohara,
  R.T., Berger, C., Ha, S.M., Rozycki, M., et~al.: Identifying the best machine
  learning algorithms for brain tumor segmentation, progression assessment, and
  overall survival prediction in the brats challenge (2019)

\bibitem{ComparativeStudy}
Baur, C., Denner, S., Wiestler, B., Navab, N., Albarqouni, S.: Autoencoders for
  unsupervised anomaly segmentation in brain mr images: A comparative study.
  Medical Image Analysis  \textbf{69},  101952 (2021).
  \doi{https://doi.org/10.1016/j.media.2020.101952}

\bibitem{SSIMAE}
Bergmann, P., Löwe, S., Fauser, M., Sattlegger, D., Steger, C.: Improving
  unsupervised defect segmentation by applying structural similarity to
  autoencoders. Proceedings of the 14th International Joint Conference on
  Computer Vision, Imaging and Computer Graphics Theory and Applications
  (2019). \doi{10.5220/0007364503720380}

\bibitem{imagenet}
Deng, J., Socher, R., Fei-Fei, L., Dong, W., Li, K., Li, L.: Imagenet: A
  large-scale hierarchical image database. In: 2009 IEEE Conference on Computer
  Vision and Pattern Recognition(CVPR). vol.~00, pp. 248--255 (06 2009).
  \doi{10.1109/CVPR.2009.5206848}

\bibitem{resnet}
He, K., Zhang, X., Ren, S., Sun, J.: Deep residual learning for image
  recognition. In: Proceedings of the IEEE Conference on Computer Vision and
  Pattern Recognition. pp. 770--778 (2016)

\bibitem{robex}
Iglesias., J.E., Liu, C.Y., Thompson, P.M., Tu, Z.: Robust brain extraction
  across datasets and comparison with publicly available methods. IEEE Trans.
  Med. Imaging  \textbf{30},  1617--1634 (2011). \doi{10.1109/TMI.2011.2138152}

\bibitem{adam}
Kingma, D.P., Ba, J.: Adam: A method for stochastic optimization (2017)

\bibitem{meissen1}
Meissen, F., Kaissis, G., Rueckert, D.: Challenging current semi-supervised
  anomaly segmentation methods for brain mri. In: Crimi, A., Bakas, S. (eds.)
  Brainlesion: Glioma, Multiple Sclerosis, Stroke and Traumatic Brain Injuries.
  pp. 63--74. Springer International Publishing, Cham (2022)

\bibitem{meissen2}
Meissen, F., Wiestler, B., Kaissis, G., Rueckert, D.: On the pitfalls of using
  the residual as anomaly score. In: Medical Imaging with Deep Learning (2022),
  \url{https://openreview.net/forum?id=ZsoHLeupa1D}

\bibitem{brats1}
{Menze}, B.H., {Jakab}, A., {Bauer}, S., {Kalpathy-Cramer}, J., {Farahani}, K.,
  {Kirby}, J., {Burren}, Y., {Porz}, N., {Slotboom}, J., {Wiest}, R., et~al.:
  The multimodal brain tumor image segmentation benchmark (brats). IEEE Trans.
  Med. Imaging  \textbf{34}(10),  1993--2024 (2015).
  \doi{10.1109/TMI.2014.2377694}

\bibitem{pytorch}
Paszke, A., Gross, S., Massa, F., Lerer, A., Bradbury, J., Chanan, G., Killeen,
  T., Lin, Z., Gimelshein, N., Antiga, L.t.: Pytorch: An imperative style,
  high-performance deep learning library. In: Wallach, H., Larochelle, H.,
  Beygelzimer, A., d\textquotesingle Alch\'{e}-Buc, F., Fox, E., Garnett, R.
  (eds.) Advances in Neural Information Processing Systems 32, pp. 8024--8035.
  Curran Associates, Inc. (2019)

\bibitem{CTAE}
Pawlowski, N., Lee, M.C., Rajchl, M., McDonagh, S., Ferrante, E., Kamnitsas,
  K., Cooke, S., Stevenson, S., Khetani, A., Newman, T., et~al.: Unsupervised
  lesion detection in brain ct using bayesian convolutional autoencoders. In:
  MIDL 2018 Conference book. MIDL (April 2018)

\bibitem{sri}
Rohlfing, T., Zahr, N.M., Sullivan, E.V., Pfefferbaum, A.: The sri24
  multichannel atlas of normal adult human brain structure. Human Brain Mapping
   \textbf{31},  798--819 (2010). \doi{10.1002/hbm.20906}

\bibitem{saase}
Saase, V., Wenz, H., Ganslandt, T., Groden, C., Maros, M.E.: Simple statistical
  methods for unsupervised brain anomaly detection on mri are competitive to
  deep learning methods (2020)

\bibitem{anogan}
Schlegl, T., Seeböck, P., Waldstein, S., Schmidt-Erfurth, U., Langs, G.:
  Unsupervised anomaly detection with generative adversarial networks to guide
  marker discovery. In: International conference on information processing in
  medical imaging. pp. 146--157 (03 2017)

\bibitem{fanogan}
Schlegl, T., Seeböck, P., Waldstein, S.M., Langs, G., Schmidt-Erfurth, U.:
  f-anogan: Fast unsupervised anomaly detection with generative adversarial
  networks. Medical Image Analysis  \textbf{54},  30--44 (2019).
  \doi{https://doi.org/10.1016/j.media.2019.01.010}

\bibitem{dfr}
Shi, Y., Yang, J., Qi, Z.: Unsupervised anomaly segmentation via deep feature
  reconstruction. Neurocomputing  (2020).
  \doi{https://doi.org/10.1016/j.neucom.2020.11.018}

\bibitem{fpi}
Tan, J., Hou, B., Batten, J., Qiu, H., Kainz, B.: Detecting outliers with
  foreign patch interpolation. arXiv preprint arXiv:2011.04197  (2020)

\bibitem{pii}
Tan, J., Hou, B., Day, T., Simpson, J., Rueckert, D., Kainz, B.: Detecting
  outliers with poisson image interpolation. In: de~Bruijne, M., Cattin, P.C.,
  Cotin, S., Padoy, N., Speidel, S., Zheng, Y., Essert, C. (eds.) Medical Image
  Computing and Computer Assisted Intervention -- MICCAI 2021. pp. 581--591.
  Springer International Publishing (2021)

\bibitem{camcan}
Taylor, J.R., Williams, N., Cusack, R., Auer, T., Shafto, M.A., Dixon, M.,
  Tyler, L.K., Henson, R.N.: The cambridge centre for ageing and neuroscience
  (cam-can) data repository: Structural and functional mri, meg, and cognitive
  data from a cross-sectional adult lifespan sample. NeuroImage  \textbf{144},
  262--269 (2017). \doi{https://doi.org/10.1016/j.neuroimage.2015.09.018}, data
  Sharing Part II

\bibitem{ssim}
Wang, Z., Bovik, A.C., Sheikh, H.R., Simoncelli, E.P.: Image quality
  assessment: from error visibility to structural similarity. IEEE transactions
  on image processing  \textbf{13}(4),  600--612 (2004)

\bibitem{BrainVAE}
Zimmerer, D., Isensee, F., Petersen, J., Kohl, S., Maier-Hein, K.: Unsupervised
  anomaly localization using variational auto-encoders. In: Medical Image
  Computing and Computer Assisted Intervention -- MICCAI 2019. pp. 289--297.
  Springer International Publishing (2019)

\bibitem{mood}
Zimmerer, D., Petersen, J., Köhler, G., Jäger, P., Full, P., Roß, T., Adler,
  T., Reinke, A., Maier-Hein, L., Maier-Hein, K.: Medical out-of-distribution
  analysis challenge (Mar 2020). \doi{10.5281/zenodo.3784230},
  \url{https://doi.org/10.5281/zenodo.3784230}

\end{thebibliography}
\end{document}